Measuring impact in research evaluations:

A thorough discussion of methods for, effects of,

and problems with impact measurements

Lutz Bornmann

Division for Science and Innovation Studies

Administrative Headquarters of the Max Planck Society

Hofgartenstr. 8,

80539 Munich, Germany.

E-mail: bornmann@gv.mpg.de

**Abstract**

Impact of science is one of the most important topics in scientometrics. Recent developments show a fundamental change in impact measurements from impact on science to impact on society. Since impact measurement is currently in a state of far reaching changes, this paper describes recent developments and facing problems in this area. For that the results of key publications (dealing with impact measurement) are discussed. The paper discusses how impact is generally measured within science and beyond (section 2), which effects impact measurements have on the science system (section 3), and which problems are associated with impact measurement (section 4). The problems associated with impact measurement constitute the focus of this paper: Science is marked by inequality, random chance, anomalies, the right to make mistakes, unpredictability, and a high significance of extreme events, which might distort impact measurements. Scientometricians as the producer of impact scores and decision makers as their consumers should be aware of these problems and should consider them in the generation and interpretation of bibliometric results, respectively.

**Key words**

Impact measurement; Citation impact; Societal impact; Altmetrics; Bibliometrics

# 1 Introduction

Governments all over the world are contemplating the question of how they should distribute public money to different areas (such as building and maintaining infrastructure, educating children and young people and protecting the natural world nationally and internationally) (Khazragui & Hudson, 2015). Distribution of money over a number of different areas always makes an issue, implicitly or explicitly, of the impact which can be achieved with the investment in any one them (Morgan, 2014). Questions arise, such as will an investment in protecting the natural world create a better environment for humans and increase species diversity? Does investment in education have a positive impact on the strength of the national economy by bringing well-educated young people and adults into the labour market? As science competes with other areas in society for public money, it is also faced with the challenge of demonstrating its value to society (Cohen et al., 2015). Basic research in particular undergoes close scrutiny for this purpose: scientists can appreciate its value to society, but politicians can hardly do so (Bornmann, 2012, 2013b).

Politicians focus on solutions to real-world issues (such as water and energy usage or the protection of the environment) and would like to know the extent to which research is contributing to the work done in these areas. Thus, they do not want to know only about the general impact of research on society, but also about the specific impact on socially relevant issues (Finkel, 2014; Thwaites, 2014). Modern society has separated into different sections (such as the economy, science and justice) which are relatively autonomous, operate as separate entities and have their own rules (Luhmann, 2012a, 2012b). That scientometrics measures the impact of research on research is in accordance with the notion of the autonomous nature of the different parts of society. Measuring the impact of research in a way that relates to specific socially relevant issues is counter to this notion of autonomy and can mean focussing on subjects about which science would normally not be concerned (because there is nothing more to research in that area, for example) (Douglas, in press).

In a number of countries (including Australia, Belgium, France, Italy, New Zealand, and the UK), national evaluation systems have become established which require science to account for funding and show that the money invested in it has not been wasted (Abramo & D'Angelo, 2011; Derrick & Pavone, 2013). These systems evaluate not only the impact of academic institutions on scientific progress, but also on the economy, environment, defence and public health. Since 2014 the UK, for example, has had the Research Excellence Framework (which replaced the Research Assessment Exercise, RAE) which evaluates the universities using peer review, case studies and metrics (King's College London and Digital Science, 2015; Thwaites, 2014). The national evaluation systems are used, on the one hand, to evaluate national research in a global context and whether impact on science and beyond has been reached. The main question is whether national research performs better or worse than the world average. The systems are used, on the other hand, to allocate research funds which are thus allocated more strictly in accordance with competition criteria than in the countries which do not have a system of this kind.

Since impact measurement of scientific outcomes is in a state of far reaching changes (Bornmann, 2014), this paper describes recent developments in this area. For that the paper discusses how impact is generally measured within science and beyond (section 2), which effects impact measurements have on the science system (section 3) and which problems are associated with impact measurement (section 4). The problems associated with impact measurement constitute the focus of this paper: It will be shown that science is marked by inequality, random chance, anomalies, the right to make mistakes, unpredictability, and a high significance of extreme events which lead to problems with impact measurements focussing on high aggregation levels and assessing continuity in research.

In the following sections it is not intended to present a complete literature overview of studies dealing with impact on science and beyond, but to discuss recent developments in impact measurements based on selected key literature. The key literature has been identified (1) in literature databases (e.g. Web of Science, Thomson Reuters, and Scopus, Elsevier) using search terms like "citation impact", "research evaluation", and "bibliometrics" and (2) in corresponding literature overviews (e.g. de Bellis, 2009; Hicks & Melkers, 2013).

## 2  Measurement of impact

Even though metrics and case studies are being used more frequently to evaluate research in various countries, the backbone of modern research evaluation has been and remains the peer review process in which colleagues mutually evaluate their research output (Bornmann, 2011b). Peer review is the oldest method of research evaluation and its use is closely associated with the development of modern science (Geisler, 2000; Virelli, 2009). It is only when research results are monitored by qualified experts in the same field to ensure that they meet certain standards that a research area can make reliable and valid statements. However, peer review comes up against its limits when a large number of research units need to be reviewed. In national evaluation systems, as a rule nearly every university and non-university research institution in a country undergoes scrutiny. To evaluate a larger number of units, quantitative methods are therefore used in addition to peer review, which apply indicators to measure the output and/or the impact of research (de Bellis, 2009; Hicks & Melkers, 2013).

The most important quantitative method is currently bibliometrics, in which citations are used to measure the impact of publications (Hicks, Wouters, Waltman, de Rijcke, & Rafols, 2015). For example, the Australian Research Council (ARC) uses mainly citation analysis to evaluate research in the natural sciences. It is only in the engineering and social sciences and in the humanities that the ARC uses the peer review of selected publications or outputs, as citation analysis is inadequate for these disciplines (Sheil, 2014). Bibliometrics is the favoured method of research evaluation primarily because papers and books are the most important products of science. Furthermore, citation counts provide information about how useful these products have been for scientists working in the same or similar disciplines (Moed, 2005).

So far, no standard method – such as bibliometrics – have emerged that can measure the benefit of research to society (that is, the broader impact of research) reliably and with validity (Campbell & Grayson, 2014; National Research Council, 2014). The search is ongoing in scientometrics for a "citation equivalent", with which "activities such as working in industry, contributions to government reports or communication of research outcomes to audiences other than a researcher's peers" (Thwaites, 2014, p. S59) can be measured. Currently, the universities mainly use case studies to determine the value of their research for society (in the UK's REF, for example). For proponents, "case studies are the only viable route to assessing impact; they offer the potential to present complex information and warn against more focus on quantitative metrics for impact case studies" (Wilsdon et al., 2015, p. 49). However, the use of case studies in evaluation is criticised because this approach can be used very selectively, if the universities report only those results with the best impact (Morgan, 2014). Furthermore, case studies are expensive and time-consuming to prepare (King's College London and Digital Science, 2015; Sheil, 2014).

The problem of finding a suitable method with which to measure societal impact also indicates that, all in all, it is much more difficult to measure societal impact than scientific impact (National Research Council, 2014). This difficulty arises primarily because there are many different target groups for the impact on society (while scientific impact is always the impact on science). Societal impact can be the impact on policy makers, business, culture or other (more specific) sections of society (Khazragui & Hudson, 2015). To measure societal impact, therefore, a decision needs to be made in each instance about what that impact actually is (Morgan, 2014). In the past, two methods of measuring societal impact based on a citation-equivalent have emerged as particularly useful: evaluating citations in patents (technological impact) (Austrian Science Fund, 2007; Kousha & Thelwall, in press) and in clinical guidelines (medical impact) (Lewison & Sullivan, 2008; Thelwall & Maflahi, 2015). Both approaches have the following advantages: (1) societal impact can be measured in a similar way to scientific impact (and sound data analysis methods – developed over decades of scientometrics research – are therefore available). (2) The fact that they are based on citations means that non-reactive, relatively objective and extensive data is available. (3) Patents and guidelines are available for the evaluation in a relatively freely accessible form and – compared with other data – can be evaluated with a reasonable amount of effort.

As well as clinical guidelines and patents, another source of data has emerged in recent years with which the impact of research in society could be measured: alternative metrics (altmetrics). "Altmetrics … is a term to describe web-based metrics for the impact of scholarly material, with an emphasis on social media outlets as sources of data" (Shema, Bar-Ilan, & Thelwall, 2014). The term was proposed by Priem, Taraborelli, Groth, and Neylon (2010). In altmetrics, counts of views, downloads, clicks, notes, saves, tweets, shares, likes, recommends, tags, posts, trackbacks, discussions, bookmarks, and comments are counted. For example, altmetrics are presented for single publications in a database such as Scopus (Elsevier), or by a publisher such as the Public Library of Science (PLOS) (Liu, Xu, Wu, Chen, & Guo, 2013; Zahedi, Costas, & Wouters, 2014). In general, altmetrics are log data which measures individual mentions of publications (such as downloads) over a certain period of time (Haustein, 2014).

The "Altmetric for Institutions" tool from the company Altmetric allows institutions to track, monitor and report on the broader impact of research (www.altmetric.com/institutions.php). It counts, analyses and, with some processing, presents online mentions of publications issued by an institution on platforms such as Mendeley,

Twitter, CiteULike and Facebook. However, Altmetric analyses more than mentions on social media platforms. Its analyses also draw on government policy documents and other sources for mentions of scholarly articles. Particularly government policy documents are an important source as the impact of research on policy making can be thus made quantifiable (Bornmann, Haunschild, & Marx, in preparation; Liu, 2014). This systematic way of measuring impact on policymaking is potentially of interest in the social sciences and the humanities in particular (Hammarfelt, 2014). In these disciplines, it is almost impossible to apply traditional bibliometrics and researchers in scientometrics are looking for alternative ways of evaluating research (Hug, Ochsner, & Daniel, 2013).

According to Thwaites (2014), health researchers are particularly interested in the impact that their research has beyond science (Thonon et al., 2015). Health researchers would like to exert impact on medical practitioners with their research. Furthermore, government is very interested in the impact of research on health care as the cost of research in this area is very high. As impact – not just in the area of health care – is often only measurable after several years, proposals have been made in societal impact research to measure the endeavours (of institutions) to achieve impact and to honour them accordingly. Instead of "monitoring impact", it would be more of a matter of "monitoring progress towards impact" (Morgan, 2014). Many Dutch organisations involved in quality assurance cooperate in a project entitled Evaluating Research in Context (ERiC), which has set itself the goal of developing methods for societal impact assessment (ERiC, 2010). One significant result of the project is that productive interaction is a necessary requirement for research to have a societal impact: "There must be some interaction between a research group and societal stakeholders" (ERiC, 2010, p. 10). Such interactions can be in the form of personal contact (e.g. joint projects or networks), publications (e.g. educational and assessment reports) (Bornmann & Marx, 2014) and artefacts (e.g. exhibitions, software or websites). All these interactions can be counted as activities to achieve societal impact.

# 3     Effects of impact measurement

Since bibliometric indicators have obtained a general acceptance in science policy and attained applied relevance in research evaluation, adaptation strategies by scientists resulting from the use of these indicators for science funding decisions have been reported (Evidence Ltd., 2007; Lawrence, 2003). Bornmann (2011a) has proposed designating these strategies "mimicry in science". Adaptation strategies can cast general doubt on the system of research evaluation: Goodhart's law states that "once a measure becomes a target it ceases to be a good measure" (McGilvray, 2014, p. S66). There is a risk that scientists are more concerned about the marketing of their research products than the content of their research. Scientists apply strategies that enable them to comply with bibliometric accountability and to secure funds for their own research. Some of these strategies are as follows:

(1) Researchers do not summarise the results of a project in a single publication, but distribute the results over many publications to create the appearance of higher productivity (Bornmann & Daniel, 2007; Mallapaty, 2014). This publication strategy is referred to as "salami slicing". However, having several papers based on the same project is not always an attempt to game the system by appearing productive, but can have substantive, methodological and practical rationales which are entirely legitimate. (2) As scientists as a rule strive to be judged as world class, they try to get their papers published in prestigious journals (Finkel, 2014). They do not choose those journals which are most suitable for their manuscripts, but those which are most highly regarded by their colleagues. In addition to these changes to publishing behaviour, there is also (3) a pressure on scientists "to change research focus to better align with mainstream or more highly esteemed fields" (McGilvray, 2014, p. S64), (4) a greater reluctance in established researchers to work with early career researchers (as they often publish in less prestigious journals), and (5) a tendency in institutions to appoint academics who already have a long publication record instead of (younger) talented researchers who do not (McGilvray, 2014).

Establishing measurement of the societal impact as part of research evaluation – alongside measuring the impact on science – would lead to scientists' angling their work more towards groups of people outside of science. For example, scientists could write reports of (own) research results at a specific group of readers outside of science (e.g. politicians). Alternative metrics could then be used – as with traditional citations – to measure the impact of the reports outside of science. If there were a desire to implement a system of output measurement (number of reports) and impact (number of mentions of the reports) like this in a national evaluation system, it would be necessary to consider in advance whether this kind of measurement would have the intended effects. (1) Is the expected adaptation in scientists' behaviour, for example that they write reports for other status groups in society in addition to producing their research papers, a desirable one? (2) Is science making a greater contribution to solving real-world issues (such as energy usage or environmental protection) with these new products, and do the economy, the environment, defence and public health in a country (see above) gain anything from it? Do the reports result in a benefit for progress in technology, health care etc. which would not be expected without the reports? According to Morgan (2014), the development of a suitable evaluation methodology, which takes account of these and similar questions is a "a big sticking point" (p. 75) which should be addressed.

## 4    Problems of impact measurement

In the following several problem areas of impact measurement are discussed. People dealing with impact data should be aware of these problems and should consider them in producing and interpreting the results.

Evaluations cost time and money, irrespective of how they are structured. A science system with evaluation should be significantly more successful than a science system without evaluation in order to be able to compensate for the disadvantage in terms of time and money. As the measurement of societal impact has arisen as a new (quantitative) element in evaluations over recent years (Ovseiko, Oancea, & Buchan, 2012), the question arises of how important it really is to include the measurement of societal impact in a national evaluation system, in terms of the balance of cost of and yield from evaluations.

Science is a part of society marked by inequality, random chance, anomalies, the right to make mistakes, unpredictability and a high significance of extreme events which are discussed in the following. As this section is intended to illustrate, these characteristics can lead to problems with measuring the impact of scientific performance which focusses on research units on a higher aggregation level and assumes continuity in research activities.

**Inequality**: The instance of inequality is very high in science. In almost every set of papers (by a researcher or an institution) there are many papers which are not cited at all or are hardly cited and only a few are highly cited papers (Seglen, 1992). As the findings of Bornmann, de Moya-Anegón, and Leydesdorff (2010) illustrate, papers contributing to scientific progress in a discipline predominantly lean on the few previously important contributions than on papers contributing little. However, it is not only on the level of individual papers, but also on the level of individual researchers that we see great inequality affecting publication output and citation impact. Ioannidis, Boyack, and Klavans (2014) investigated 15,153,100 publishing scientists (distinct author identifiers) in the Scopus database. Their results show that "only 150,608 (<1%) of them have published something in each and every year in this 16-year period (uninterrupted, continuous presence [UCP] in the literature). This small core of scientists with UCP are far more cited than others, and they account for 41.7% of all papers in the same period and 87.1% of all papers with >1000 citations in the same period".

**The right to make mistakes and the significance of extreme events**: Science is characterised not only by skewed distribution, but also by the right to make mistakes.

According to Popper (1961), scientific issues, hypotheses and problems are solved through trial and error (and not with the empirical confirmation of previously formulated hypotheses). When a problem is examined by researchers, the attempts at solution are tested empirically, whereby the poor alternatives are identified and removed. Scientific research is therefore always prone to error and associated with the right to use fallible trial and error, to take risks and unrecognised (unorthodox or intuitive) routes (Bornmann, 2013a). Scientific progress based on trial and error is not however cumulative; i.e. the ongoing production of knowledge units, but happens through extreme events which Kuhn (1962) has called scientific revolutions. Important publications in a discipline lead to completely new thinking, which is reflected by changes in the taxonomy used (Wray, 2011). The taxonomy before a revolution is fundamentally different from the taxonomy after the revolution.

**Anomalies**: Where inequality, error and extreme events are important components in a system, we can assume that anomalies also play a key role. As the studies by Bornmann et al. (2010) and Ioannidis et al. (2014) have shown, science is determined essentially by a few elements (such as publications and scientists) and not by a mass of many. Even the journals which are continuously analysed for the literature database Web of Science (Thomson Reuters) are selected on the assumption that scientific progress can be represented by a few core journals and therefore the vast majority of journals need not be included (Garfield, 2006). As bibliometric analyses are as a rule undertaken on a higher aggregate level (such as institutions or countries), the effect of anomalies (the few highly cited papers) on the whole is overseen or they are treated as problems yet. For example, Göttingen University only achieved a good position in a former release of the Leiden ranking (which used a mean-citation-based indicator to rank institutions) because it could publish <u>one</u> extremely highly cited publication (Waltman et al., 2012). With analyses on a higher aggregate level, where data is very skewed, a few anomalies are responsible for the results but their immense influence on the overall result (such as the average citation impact of an institution) is frequently not visible. An institution does not acquire an excellent score in an impact measurement due to the majority of scientists it employs and their publications, but due to its few very successful scientists or their small number of highly cited papers.

**Randomness and unpredictability**: Campanario (1996) has published some good evidence for the existence of serendipity in science: He examined 205 Citation Classics commentaries from highly-cited papers in the recent history of science. "Authors of 17 Citation Classics commentaries (8.3%) mention some kind of serendipity in performing the research reported in the highly cited paper." The term 'serendipity' is closely associated with the name of Robert K. Merton who published a book on this topic with Elinor Barber (Merton & Barber, 2004). Much that is of importance in science is the result of happy coincidence – arising from unknown causes or causes which are almost impossible to find rationally. The discovery of X-rays and penicillin are good examples of this (Ban, 2006). The existence (and importance) of random elements in scientific work indicate that there are obviously parallels between the acquisition of knowledge in science and evolutionary processes in nature (Feist, 2006; Gieryn, 1995). While living beings adapt better with random genetic changes in nature, scientific findings are the result (among other things) of studies which are by chance appropriate. Researchers find things that they were not even looking for.

It follows from these random elements in the process of creating knowledge that important progress in science is often unpredictable. There are many examples in science where the importance of certain research results has appeared only decades after their publication (Ke, Ferrara, Radicchi, & Flammini, 2015; van Raan, 2004). At the time of publication, only a few scientists (at least the reviewers of the manuscript) were expecting that it would have any significance for the scientific community. For example, the "Shockley-Queisser limit" (Shockley & Queisser, 1961) describes the limited efficiency of solar cells on the basis of absorption and reemission processes (Marx, 2014). The original reception of the paper in terms of citations was initially hesitant. However, the paper has become one of the highly cited papers in its field. The citations of the paper published by Shockley and Queisser (1961) developed relatively synchronously with the rapidly growing solar cell and photovoltaic research area.

However, even when results have been presented in a discipline and only later proved to be particularly important, experts in the same field frequently do not recognise their wider significance and therefore do not cite these papers. Marx and Bornmann (2010, 2013) showed in two bibliometric studies on scientific revolutions that a number of publications, which turned out to be important for a revolution, were rarely cited. An important discovery is therefore often credited to the author who links the various strands of knowledge required for the discovery in a meaningful way and not those who have contributed or published a crucial empirical result or theory. Bornmann and Marx (2012) have proposed the designation "Anna Karenina Principle" for this process where different and necessary strands of knowledge derived from theoretical and empirical processes are brought together to create a revolutionary discovery.

The importance of inequality, anomalies, randomness, unpredictability, error and revolutionary events in the process of scientific discovery is frequently seen in literature as evidence that science cannot be planned, managed or measured and therefore it would be superfluous to measure impact within the framework of research evaluations (Schatz, 2014). There are especially two features of evaluations which lead to problems with impact measurements of scientific activities operating against the background of inequality, anomalies, randomness, unpredictability, error and revolutionary events. These features are (1) the units usually evaluated in a measurement of impact, and (2) the continuity desired in research performance:

**Evaluated units**: The impact measured is frequently that of institutions (such as universities) or countries (such as the BRICS states). University rankings such as the Times Higher Education World University Rankings (www.timeshighereducation.co.uk), which report regularly on the performance of universities, are a good example (Hazelkorn, 2011). As the results of Bornmann, Mutz, and Daniel (2013) based on the universities considered in the Leiden Ranking show, only 4.3% of the variance in institutional citation impact can be attributed to differences among the universities. The rest, namely 95.7%, is allocated to variance within the universities (that is, to departments, research groups, and individual researchers). The heterogeneity of research performance which is the result of inequality, anomalies, randomness, unpredictability, error and revolutionary events is therefore so pronounced within the universities that it seems questionable to measure impact at institutional level. Instead, it seems more appropriate to evaluate research performance on the basis of smaller units which are responsible for the important results leading to significant scientific progress.

A possible reason for the popularity of impact analyses at institutional and country level is undoubtedly the availability of data. It is much harder to collect data on research groups and individual scientists than on institutions and countries. The data on individuals is difficult to identify because different people share the same names (Boyack, Klavans, Sorensen, & Ioannidis, 2013). While authors of papers always give the name of the institution where they work and the country where the institution is located, they rarely provide any information about the research group, the department or similar. Another reason for the popularity of institutions and country analyses is undoubtedly the political focus on these units (Hazelkorn, 2011). The government of a country sees itself in competition with other countries and therefore calls for the relevant country-based studies (National Science Board, 2014).

**Continuity**: Generally speaking, the evaluation of science is continuous. Institutional evaluations are repeated every few years in order to assess whether the steady input of financial resources will create a steady output and impact. Each new evaluation of an institution is expected to show that the output and impact of the previous evaluation has been at least achieved, and preferably surpassed. Scientists at an institution are therefore asked to produce a constant and increasing flow of important research results which can be published in high-impact journals and later achieve a high impact. However, the features of science described above (inequality, anomalies, randomness, unpredictability, error and revolutionary events) do not give reasons to expect a linear relationship between input and output in scientific work. If the relationship is assumed to be linear, there is the danger that the disparity between the aspiration of the evaluation and the reality of the research process results in scientific misconduct to adjust the reality to the aspiration (Bornmann, 2013a). To counter the dissatisfaction at this disparity, misconduct is taken into account for, or in, an evaluation.

Merton's (1938) general theory on misbehaviour in society can be used to explain the connection between dissatisfaction and aberrant behaviour in science. Merton (1938) distinguishes between three factors in social structure to explain aberrant behaviour in society (such as criminality in the USA): (1) Certain wishes and expectations represent important cultural goals in a society; (2) norms which state rules for how these goals should be reached and (3) the distribution of resources required to reach the goals. For Merton (1938), aberrant behaviour results when a social structure makes it impossible to achieve the cultural goals shared and internalised by a society (in the USA, this would be individual wealth, for example) with socially accepted means (such as honest work). Aberrant behaviour is likely to occur in persons in a certain group if they can only achieve the goals dictated by society with great difficulty if they use legitimate means. It arises in a situation in which certain symbols for success are strongly overemphasised by the group, but only a small part of these symbols can be acquired by legitimate means, due to the way that resources are distributed.

Against the background of this general mechanism to explain aberrant behaviour, scientific misconduct would result from a situation in which the researchers are faced with the goals set up by the continuous evaluations ('winning the game') which can only be reached with great difficulty, or not at all, within the cognitive and social norms in science ('winning through circumscribed modes of activity'). The research results are falsified or massaged in order to produce for the evaluations new results which are published in reputable journals. McGilvray (2014) describes an attempt at gaming in a national evaluation system: "Perhaps the most egregious example of an attempt to game the system occurred in New Zealand in 2006. One leading university reclassified dozens of staff members, notably those who were PBRF [Performance-Based Research Fund] - eligible but performed little active research. By reclassifying inactive researchers away from subjects such as economics and biology to fields such as philosophy and religious studies, the university would improve its standing in the former fields. The surge in the number of New Zealand philosophers piqued the curiosity of PBRF reviewers who eventually reversed the classifications" (p. 66).

The recourse to aberrant behaviour (such as fabricating and falsifying research results and reassigning people for an evaluation) can be attributed to (1) excessive emphasis on objectives in the continuous evaluations, which are defined by certain symbols of success in research (such as higher citation impact); (2) a generally diminished importance of rules intended to be applied to the process of achieving the objectives (such as achieving a high citation impact) and (3) the restricted availability of financial and personnel resources with which to achieve the objectives.

## 5   Conclusions

There is no doubt that there would be no progress in society without science. Scientific research results in new technology, a greater understanding of our planet and the universe and in treatments which allow us to live longer and more healthily (Campbell & Grayson, 2014). Scientists are investigating the reasons for climate change and are working on more secure and sustainable solutions to provide us with energy. They have significantly improved the accuracy of weather forecasts and made a substantial contribution to the control of infectious diseases. Despite these achievements, today's audit society (Dahler-Larsen, 2011; Power, 1999) expects that science and other areas of society are accountable to the state for their outcomes and that they address the specific problems which governments believe confront society with greater dedication. As these specific demands on science represent a relatively new phenomenon (arising in recent decades), there is a question of how they are changing or have changed science. In the view of Luhmann (2012a, 2012b), we can expect that science will take on these challenges and apply itself more to them; however, this will only be possible with the methods, instruments and practices with which science has always operated. Research as an activity would not, therefore, in the view of Luhmann (2012a, 2012b) change fundamentally.

Even if the way in which new findings are made probably does not change, it is likely that the way they are published will. If the impact of research not only on science but more widely across various areas of society is measured, it would be advantageous to formulate the findings for the relevant readership. Appropriately formulated texts will have a better chance of creating an impact than typical scientific texts (that is, papers published in scientific journals). This preparation could be undertaken by others, such as science journalists. However, journalists are at home in their own sub-section of society, the mass media (Luhmann, 2000), and operate according to the rules which apply there in order to reach a wider audience. If this preparation were to be undertaken by the scientists themselves, their publishing habits would have to change (Bornmann & Marx, 2014).

This paper has examined the measurement of impact and investigated the reasons why the impact of science is measured in a certain way. It has focussed on the problems of impact measurement and its unintended effects. Both areas should be accorded particular attention nowadays as impact measurement is undertaken on a wider scale than some years ago. In-depth scientometric research is required to define and assess possible metrics-based evaluation systems sensibly, effectively and systematically. However, this research should not only look at the development of reliable and valid indicators, but also at the effects and problems produced by these systems. A key question should be in how far these systems can really result in science performing better. In these studies, for example, the performance of different countries could be compared which use national evaluation systems or not. The challenge in conducting these studies will be to separate the effect of the evaluation system on a country's performance from other effects (e.g. the availability of more or less money for research from the state).